\documentclass[fleqn,usenatbib]{mnras}

\usepackage{newtxtext,newtxmath}
\usepackage{cancel}

\usepackage[T1]{fontenc}

\DeclareRobustCommand{\VAN}[3]{#2}
\let\VANthebibliography\thebibliography
\def\thebibliography{\DeclareRobustCommand{\VAN}[3]{##3}\VANthebibliography}

\usepackage{graphicx}	% Including figure files
\usepackage{amsmath}	% Advanced maths commands
\usepackage{bm}	% Bold math symbols
\usepackage{tikz}
\usepackage{soul}
\usepackage{xcolor}
\newcommand{\res}[1]{\textcolor{black}{#1}}
\title[PTA-GSS: Marginal likelihood estimation]{Generalized Steppingstone Sampling: Efficient marginal likelihood estimation in gravitational wave analysis of Pulsar Timing Array data}

\author[EMZ et al.]{
El Mehdi Zahraoui,$^{1}$\thanks{E-mail: elmehdi.zahraoui@autuni.ac.nz}
Patricio Maturana-Russel,$^{1,3}$
Willem van Straten,$^{2}$
Renate Meyer$^{3}$ 
 and Sergei Gulyaev$^1$
\\
$^{1}$Department of Mathematical Sciences, Auckland University of Technology, Private Bag 92006, Auckland 1142, New Zealand
\\
$^{2}$Manly Astrophysics, 15/41-42 East Esplanade, Manly, NSW 2095, Australia
\\
$^{3}$Department of Statistics, University of Auckland, Auckland 1142, New Zealand
}

\date{Accepted XXX. Received YYY; in original form ZZZ}

\pubyear{\the\year{}}

\begin{document}
\label{firstpage}
\pagerange{\pageref{firstpage}--\pageref{lastpage}}
\maketitle

\begin{abstract}

Globally, Pulsar Timing Array (PTA) experiments have revealed evidence supporting an existing gravitational wave background (GWB) signal in the PTA data set. Apart from acquiring more observations, the sensitivity of PTA experiments can be increased by improving the accuracy of the noise modeling. In PTA data analysis, noise modeling is conducted primarily using Bayesian statistics, relying on the marginal likelihood and \res{the} Bayes factor to assess \res{the} evidence. We introduce generalized steppingstone (GSS) as an efficient and accurate marginal likelihood estimation method for the PTA-Bayesian framework. This method enables \res{low-cost} estimates with high accuracy, especially when comparing expensive models such as the Hellings-Downs (HD) model or the overlap reduction function model (ORF). We demonstrate the efficiency and the accuracy of GSS for model selection and evidence calculation by reevaluating the evidence of previous analyses from the North American Nanohertz Observatory for Gravitational Waves (NANOGrav) 15 yr data set and the European PTA (EPTA) second data release. We find similar evidence for the GWB compared to the one reported by the NANOGrav 15-year data set. Compared to the evidence reported for the EPTA second data release, we find a substantial increase in evidence supporting GWB across all data sets.
\end{abstract}


%%%%%%%%%%%%%%%%%%%%%%%%%%%%%%%%%%%%%%%%%%%%%%%%%%

%%%%%%%%%%%%%%%%% BODY OF PAPER %%%%%%%%%%%%%%%%%%

\section{Introduction}

The pulsar timing array (PTA) emerged from the concept of a single pulsar-Earth baseline to observe gravitational waves (GW) in the low-frequency nanohertz band \citep{sazhin1978opportunities}. The PTA experiments were designed as a network of pulsars that could unveil the influence of GWs on the pulsating radio beams emitted from pulsars in the form of quadrupolar correlations \citep{hellings1983upper}. Millisecond pulsars (MSPs) became the key to achieving this objective due to their long-term stability, rivaling Earth's atomic clocks \citep{matsakis1997statistic}. Since the initial MSP timing analysis \citep{stinebring1990cosmic}, the red noise (RN) in the MSPs' dataset has pointed to a promising future as observational radio astronomy and computational capability constantly improve. The next step for the PTA community focused on gathering more data and refining the pulsar timing model to test and derive a realistic sensitivity threshold to observe GWs \citep{verbiest2009timing,cordes2010measurement,hazboun2019realistic}. Recently, major PTA collaborations such as the North American Nanohertz Observatory for Gravitational Waves \cite[NANOGrav;][]{McLaughlin_2013}, the Australian Parkes PTA  \cite[PPTA;][]{manchester2013parkes}, the European PTA  \cite[EPTA;][]{ferdman2010european}, the Indian PTA  \cite[InPTA;][]{joshi2018precision}, reported positive statistical support for gravitational wave background (GWB) in the latest data releases \citep{reardon2023search,agazie2023nanograv,antoniadis2023second}.

The PTA community has established a rigorous PTA-Bayesian framework \citep{van2009measuring}, consistently expanding and acquiring new numerical and statistical techniques \citep{lentati2014temponest,johnson2023nanograv}. This framework has enabled a simultaneous fit for different noise models and GWB signals while having the ability to compare the evidence of each different model \citep{van2014new}. Before evaluating the GWB evidence, it was proven that a common red signal (CRS) is evident across different PTA datasets: NANOGrav \citep{arzoumanian2020nanograv}, EPTA \citep{chen2021common}, and PPTA \citep{goncharov2022consistency}. These findings resulted in a surge of interest in defining the nature of these CRSs and potentially revealing the buried GWB signal. Through this process, an extensive number of models were used to mitigate different sources of red noise that can affect the PTA's sensitivity either from pulsar contribution such as pulsar spin noise \citep{verbiest2009timing} and dispersion measure (DM) variations \citep{jones2017nanograv}, or other noise sources such as spatial correlations due to the solar system ephemeris and the atomic clocks used in this experiment \citep{tiburzi2016study}. To test and account for all noise models, the full PTA-Bayesian analysis results in a large model space. Down the line, incorporating new MSP datasets and updated noise models will further expand this space significantly.

In the Bayesian framework, hypotheses comparison is done by evaluating the \textit{Bayes factor}, which is the ratio of \textit{marginal likelihoods} of different models. This criterion is \res{used} to assess the evidence for each hypothesis \citep{newton1994approximate}. For the PTA, model selection and evidence estimation are currently carried through various Bayesian techniques \citep{agazie2023nanograv,antoniadis2023second}. Among these methods, thermodynamic integration \citep[TI;][]{lartillot2006computing} stands out as the cornerstone method for marginal likelihood estimation. In parallel, nested sampling \citep{skilling2006nested} is implemented to provide not only estimates of marginal likelihood but also to generate posterior samples. Additionally, reweighting \citep{hourihane2023accurate} was defined to estimate the marginal likelihood for an expensive model using an approximately matching model. Finally, the Hypermodel method, implemented using the Savage-Dickey density ratio approximation as a form of product space sampling for nested models \citep{carlin1995bayesian,lodewyckx2011tutorial}, is routinely used across the whole PTA analyses. \cite{johnson2023nanograv} discuss these methods in the context of the PTA, where nested sampling is used for models with small parameter spaces. In this case, nested sampling yields less accurate estimates with a similar cost to the Hypermodel method. In parallel, TI estimates become expensive for large PTA models and can lead to biased estimates if the required computational power is lacking. Consequently, TI is employed only where the Hypermodel fails to provide evidence estimates. Although the Hypermodel method was established as the predominant method, it is limited to nested models and requires assigning arbitrary weights to each model to avoid the Markov chain getting trapped in a particular model. Alternatively, we introduce generalized steppingstone sampling \citep[GSS;][]{fan2011choosing} as a new method to GW astronomy and in particular the PTA-Bayesian framework. GSS is intended to bypass most of the current issues with marginal likelihood estimation and provide an accurate marginal likelihood estimate at a \res{lower  computational price}. 

This article is structured as follows. Section~\ref{marginal likelihood estimation} outlines model selection through the Bayes factor and briefly defines the TI, SS, and GSS methods. Section~\ref{simulation study} demonstrates a comparison between these methods' performance applied to a Gaussian model. Thereafter, section~\ref{Application to PTA} first discusses the implementation of GSS within the PTA framework. Then, our application to different PTA analyses will revisit the GWB evidence and will demonstrate the benefits of marginal likelihood estimation in making a robust model selection and monitoring the evidence on different CRS models. Finally, section~\ref{discussion} will discuss the prospects of GSS in the PTA-Bayesian framework and the significance of different PTA application results.

\section{Marginal likelihood estimation}
\label{marginal likelihood estimation}
%\pmr{This sentence is not clear, frequentist stat does the same, no?}
%Bayesian inference examines different hypotheses to explain and model a data set while evaluating the fitting performance. 
 %\pmr{why italic?}
 %\pmr{what is the criterion for the italic?}
The Bayesian framework is based on \textit{Bayes'} theorem, which is given by
\begin{equation}
p(\boldsymbol\theta|X,M)=\frac{L(X|\boldsymbol\theta,M)\pi(\boldsymbol\theta|M)}{z(X|M)}
    \label{bayes}
\end{equation}
for a model $M$  and dataset $X$, where $\boldsymbol\theta\in\Theta$ is the parameter vector, $p(\boldsymbol\theta|X,M)$ the posterior probability density of $\boldsymbol\theta$, $L(X|\boldsymbol\theta,M)$ the likelihood function  and $\pi(\boldsymbol\theta|M)$ the prior density. The constant $z(X|M)$, commonly denoted as $z$, is the \textit{marginal likelihood}. The marginal likelihood is a multi-dimensional integral over the parameter space $\Theta$ defined by 
\begin{equation}    \textit{z}=\int_{\Theta}L(X|\boldsymbol\theta,M)\pi(\boldsymbol\theta|M)\text{d}\boldsymbol\theta,
    \label{marginal_like}
\end{equation}
where  $\pi(\boldsymbol\theta|M)$ is a proper prior, i.e.,  $\int_{\Theta}\pi(\boldsymbol\theta|M)\text{d}\boldsymbol\theta=1$. The integral $z$ is a prior weighted average of the likelihood which quantifies the goodness of fit of model $M$ to data $X$. For a set of two models $M_1$ and $M_2$ fitted to a dataset $X$, the marginal likelihoods $z(X|M_{1})$ and $z(X|M_{2})$ can be used to compare these models such that the model with the higher marginal likelihood is preferred. Hence, the \textit{Bayes factor} (BF), defined by
\begin{equation}
    \text{BF}_{(M_{1} / M_{2})}=\frac{z(X|M_{1})}{z(X|M_{2})},
    \label{bayes_factor}
\end{equation}
 is used as a Bayesian criterion to compare $M_{1}$ to $M_{2}$ and perform model selection. Specifically, the \textit{Bayes factor} is equivalent to the ratio of the posterior distributions of the models when the prior probabilities for the models are equal. In contrast, alternative performance tests are used to compare models such as the Bayesian information criterion (BIC) and the Akaike information criterion (AIC). Although the BIC and the AIC penalize unnecessary complexity by applying a uniform penalty for each parameter, these approaches ignore the importance of priors in the Bayesian analysis. The Bayes factor allows for tailored penalties based on the chosen priors which helps to ensure that overly complex models are penalized appropriately, reducing the risk of overfitting.
 
Apart from simple conjugate cases, the marginal likelihood has no analytical solution. In practice, considerable efforts have been made to numerically approximate $z$  through Markov chain Monte Carlo (MCMC) methods such as the harmonic mean \citep{newton1994approximate}, bridge sampling \citep{meng1996simulating}, and nested sampling \citep{skilling2006nested}. Path sampling \citep{gelman1998simulating} or thermodynamic integration \citep{lartillot2006computing} was the cornerstone technique of what would be later known as the method of power posterior \citep{friel2008marginal}.  This method defines a \res{geometric} path that connects the prior with the posterior through the power posterior densities
\begin{equation}    p_\beta(\boldsymbol\theta|X,M)=\frac{L(X|\boldsymbol\theta,M)^\beta\pi(\boldsymbol\theta|M)}{z_\beta}\quad\text{with}\quad 0\leq\beta\leq1,
    \label{ppost}
\end{equation}
where $\beta$ is the inverse temperature for the power posterior density $p_\beta$. When $\beta=0$, the density $p_\beta$  is equal to the \res{proper} prior distribution\res{, which gives} $z_0=1$. When $\beta=1$, the density $p_\beta$  is equal to the posterior distribution resulting in $z_1=z$. Correspondingly, thermodynamic integration relies on the identity
\begin{equation}
    \log(z)= \int_0^1 E_{p_{\beta}}[\log(L(X|\boldsymbol\theta,M))]\text{d}\beta,
    \label{logz}
\end{equation}
where $E_{p_{\beta}}$ denotes the expected value with respect to $p_{\beta}$.  For a sequence of $\beta$ values, samples from each power posterior can be used to estimate the expected values by the sample averages.  Thus, the integral in Equation \eqref{logz} can be approximated using any technique for numerical integration, e.g., the trapezoidal rule. Thermodynamic integration can yield a good marginal likelihood estimate; nevertheless, it requires a high computational expense to produce a single estimate. This computational expense obliged  modifications to the method to reduce the estimation cost. Steppingstone sampling was proposed as an attempt to reduce TI's cost, which resulted in a later far more efficient and accurate GSS method, which is considered one of the most accurate marginal likelihood estimation methods.

\subsection{Steppingstone Sampling}
The approximation of a continuous integral induces a discretization bias in TI leading to an inaccurate estimate of the marginal likelihood. Hence, steppingstone sampling was first proposed in the phylogenetics field \citep{xie2011improving} and later introduced to LIGO GW analysis \citep{maturana2019stepping} as an alternative method to enable unbiased \res{$z$} estimates and improve model selection. \res{The SS method considers an unnormalized version of the power posterior density, given in \eqref{ppost}, defined by
\begin{equation}
\label{eq:SS q_beta}
q_{\beta}=L(X|\boldsymbol\theta,M)^{\beta}\pi(\boldsymbol\theta|M). 
\end{equation}
This density is normalized by $z_{\beta}$ yielding a normalized power posterior density
\begin{equation}
p_{\beta}=\frac{q_\beta}{z_\beta} \text{,  with 
  }\,z_{\beta}=\int_{\Theta}q_{\beta}\text{d}\boldsymbol{\theta}.
  \label{SS p_b}
\end{equation}
}\res{Note that $z_0$ represents the normalizing constant of a proper prior, which is equal to 1, whereas $z_1$ is the normalizing constant of the posterior, i.e., the marginal likelihood.  Taking this into account, SS redefines the marginal likelihood as} a product of $K-1$ intermediate ratios given by 
\begin{equation}
    z=\frac{z_1}{z_0}=\frac{z_{\beta_1}}{z_{\beta_0}}\frac{z_{\beta_2}}{z_{\beta_1}}...\frac{z_{\beta_{K-1}}}{z_{\beta_{K-2}}}= \prod_{k=1}^{K-1}\frac{z_{\beta_{k}}}{z_{\beta_{k-1}}},
    \label{z_SS}
\end{equation}
where $\beta_{0}=0<\cdots<\beta_{k-1}<\beta_k<\cdots<\beta_{K-1}=1$. Importance sampling can be used to estimate each of the ratios $r_{k}=z_{\beta_{k}}/z_{\beta_{k-1}}$, where $p_{\beta_{k-1}}$ acts as an excellent importance sampling distribution \res{when} $p_{\beta_{k-1}}$ is \res{similar} to $p_{\beta_{k}}$\res{, which occurs} for \res{a sufficient} large $K$\res{.}\res{ It can be shown that} the ratio\res{s} $r_k$ can be \res{expressed} as
\res{
\begin{equation}
    r_k=E_{p_{\beta_{k-1}}}\left[\left(L(X|\boldsymbol\theta,M)\pi(\boldsymbol\theta|M)\right)^{\beta_{k}-\beta_{k-1}}\right].
    \label{r_k_SS}
\end{equation}}
In practice, \res{these} ratios are estimated using \res{an} unbiased Monte Carlo \res{(MC)} estimator \res{for each ratio $r_{k}$, given by}
\res{
\begin{equation}
   \widehat{r}_k=\frac{1}{n}\sum_{i=1}^{n}L(X|\boldsymbol\theta_{k-1,i},M)^{\beta_{k}-\beta_{k-1}}.
    \label{r_hat_SS}
\end{equation}}Therefore\res{,} the \res{SS} estimate of the marginal likelihood is calculated by 
\begin{equation}
    \widehat{z}= \prod_{k=1}^{K-1}\widehat{r}_{k}=\prod_{k=1}^{K-1}\frac{1}{n}\sum_{i=1}^{n}L(X|\boldsymbol\theta_{k-1,i},M)^{\beta_{k}-\beta_{k-1}},
    \label{log_z_SS}
\end{equation}
with $\boldsymbol\theta_{k-1,i}$ samples drawn from $p_{\beta_{k-1}}$ and $n$ the number of drawn samples. To avoid numerical errors, it is preferable to work with $\log \widehat{z}$. However, this transformation introduces a bias to the $\widehat{z}$ estimate which can be overcome by increasing $K$, the number of $\beta$ chains \res{that corresponds to $n$ samples from $p_{\beta}(\boldsymbol{\theta}|X,M)$} \citep{xie2011improving}. The choice of the $\beta$ values and their dispersion can enhance the efficiency of SS and even more \res{for} TI \citep{maturana2019stepping,xie2011improving}. \res{ For $\beta_{k}$ defined as 
\begin{equation}
    \beta_{k}=\frac{k}{K}100\%\text{ quantile of Beta($\alpha,1$)}
\end{equation}
for $ k=1,\dots,K-1$}, it has been shown that the efficiency can be increased with $\alpha = 0.3$ for both methods \citep{xie2011improving}. \res{ In general, the prior is more diffuse than the posterior, leading to much greater differences between consecutive power posterior densities near the prior, i.e., for $\beta$ close to 0. Consequently, the SS method exhibits greater differences between consecutive ratios $\widehat{r}_k$ when $\beta$ is close to 0.  As a result, more power posterior densities are required near the prior, meaning that more $\beta$ values close to 0 are needed, which is effectively achieved by the quantiles of the Beta(0.3, 1) distribution.}

%$\alpha \in[0.2,0.4]$ \pmr{I'd just mention.  When they talk about that interval, they refer to a particular example} 

\subsection{Generalized Steppingstone Sampling}

The performance of the steppingstone sampling reinforced the interest in decreasing the cost of marginal likelihood estimation while maintaining high accuracy. The generalized SS was proposed by \cite{fan2011choosing} to achieve this objective. The GSS method \res{introduces an additional} distribution, referred to as the reference distribution $\pi_0$, \res{ which should be an approximation of the posterior. This constrains the parameter space explored by the MCMC method, making GSS more efficient than SS. }%\res{which helps in reducing the difference between consecutive power posteriors, leading the successive ratios $\widehat{r}_{k}$ to be more similar}.
In practice, the reference distribution can be defined as the product of convenient distributions \res{tuned by} posterior samples. \res{GSS introduces the reference distribution $\pi_{0}$ in the unnormalized posterior given in~(\ref{eq:SS q_beta}), redefining it as follows}
\begin{equation}
    q_{\beta}=[L(X|\boldsymbol\theta,M)\pi(\boldsymbol\theta|M)]^{\beta}[\pi_{0}(\boldsymbol\theta|M)]^{1-\beta}.
    \label{GSS q_b}
\end{equation}
\res{Thus, the power posterior in \eqref{SS p_b} defines a path between the reference distribution and the posterior. Note that for $\beta=0$, $p_{\beta}$ is equal to the reference distribution, while for $\beta=1$, $p_{\beta}$ is equal to the posterior.}
\res{The product of ratios that defines the SS estimator in~\eqref{z_SS} remains the same in the GSS estimator; however, it can be shown that the ratios are now defined as follows}
\begin{equation}
    r_{k}=E_{p_{\beta_{k-1}}}\left[\left(\frac{L(X|\boldsymbol\theta,M)\pi(\boldsymbol\theta|M)}{\pi_{0}(\boldsymbol\theta|M)}\right)^{\beta_{k}-\beta_{k-1}}\right].
    \label{gss r_k}
\end{equation}
The MC estimator of the ratio $r_k$ is given by
\res{
\begin{equation}
    \begin{split}
    \widehat{r}_{k}=\frac{1}{n}\sum^{n}_{i=1}\left(\frac{L(X|\boldsymbol{\theta}_{k-1,i},M)\pi(\boldsymbol{\theta}_{k-1,i}|M)}{\pi_{0}(\boldsymbol{\theta}_{k-1,i}|M)}\right)^{\beta_{k}-\beta_{k-1}},
    \end{split}
    \label{gss mc r_k}
\end{equation}}
with $n$ samples $\boldsymbol{\theta}_{k-1,i}$ \res{drawn} from $p_{\beta_{k-1}}$\res{.}
Finally, the marginal likelihood is estimated by \res{multiplying} over the \res{estimated} ratios \res{as follows} 
\res{
\begin{equation}
    \widehat{z}= \prod^{K-1}_{k=1}\widehat{r}_k=\prod_{k=1}^{K-1}\frac{1}{n}\sum_{i=1}^{n}\left(\frac{L(X|\boldsymbol{\theta}_{k-1,i},M)\pi(\boldsymbol{\theta}_{k-1,i}|M)}{\pi_{0}(\boldsymbol{\theta}_{k-1,i}|M)}\right)^{\beta_{k}-\beta_{k-1}}.
    \label{gss z}
\end{equation}}
\res{The MC estimator for each ratio $r_k$ provides an unbiased estimate of the ratio. However, this does not imply that their product is unbiased unless the estimated ratios are independent. In practice, the interaction between Markov chains in the parallel tempering algorithm in the GSS case is unnecessary due to prior knowledge of the posterior distribution. As a result, independent Markov chains can be generated, leading to independent estimated ratios and thereby eliminating bias in the GSS estimate.}

\res{The uncertainty of the GSS estimate can be approximated using the delta method--as proposed in the SS case \citep{xie2011improving}-- which relies on the assumption that the samples are independent.  In practice, this assumption might not be met, resulting in a biased uncertainty estimate.  Given that independent GSS estimates can be obtained at a relatively low computational cost, empirical uncertainty can be directly estimated. This method is adopted in the following applications.}

GSS inherits the benefits of the SS method and \res{is equal} to SS if $\pi_0(\boldsymbol\theta|M)=\pi(\boldsymbol\theta|M)$. For GSS to be efficient, one can tailor a reference distribution $\pi_0$ that approximates the posterior as the product of independent distributions\res{, each reflecting} the \res{marginal} distribution of each parameter $\theta$\res{$_j$, which is an element of the parameter vector $\bm{\theta}$}. \res{If posterior samples are available, these samples} can be used to calibrate each \res{marginal} probability density \res{of the reference} distribution \res{for each} $\theta_j$, where each distribution can be set\res{,} for example\res{,} to \res{$\mathcal{N}(\bar{\theta}_{j},S_j^{2})$,} with \res{$\bar{\theta}_{j}$ and $S_j$ being} the mean and standard deviation of these samples\res{, respectively}. 

\res{The GSS method bears similarity to the recently proposed weighting method \citep{hourihane2023accurate}.} 
It can be shown that for a target model $T$ and an approximate model $A$ with equal parameter space, the \res{r}eweighting method is a particular case of the GSS method. In this case, the reference distribution $\pi_{0}$ is equal to $\textit{likelihood}\times\textit{prior}$ of the approximate model $A$ with $K=2$. The \res{r}eweighting method is particularly useful \res{for the PTA models} when the sampling from model T is computationally expensive  \res{compared to the approximate model A}.

\section{Simulation study}
\label{simulation study}
In this simulation study, we will implement the GSS method and compare its performance to that of  SS and TI. Since the marginal likelihood $z$ of real cases generally does not have an analytical solution, we resort to a simple Gaussian model where the marginal likelihood is known \citep{lartillot2006computing}. This Gaussian model is parameterized by the vector $\boldsymbol\theta=(\theta_1,\theta_2,...,\theta_d)$ \res{with a null data vector $X$}, where the prior on $\boldsymbol\theta$ is a product of standard normal distributions $\theta_j\sim \mathcal{N}(0,\,1)$, and the likelihood is defined by
\begin{equation}
    L(\res{X|}\boldsymbol{\theta})=\prod_{j=1}^{d} \exp\left({-\frac{\theta_{j}^{2}}{2v}} \right),
    \label{simul}
\end{equation}
with $d$ the number of dimensions, and \res{a known variance $v$}.  This setup enables access to the analytical form for each component of the Bayesian model. The power posterior density can be expressed as a product of $d$ normal distribution $\mathcal{N}(0,\,v/(v+\beta))$. The density $p_\beta$ is equal to the posterior distribution when $\beta=1$ yielding a marginal likelihood equal to 
\begin{equation}
    z=\left(\frac{v}{1+v}\right)^{d/2}.
    \label{simul_z}
\end{equation}
Since the analytical power posteriors are available, we sample directly from these densities to avoid MCMC sampling. For SS and TI, $n$ independent samples are generated for each $\beta_{k}$ chain, and then $K$ the number of chains is increased until the numerical estimate reaches the true value. In contrast, the GSS estimator is executed in two steps. First, independent samples from the posterior distribution (calibration samples $N_{\text{cal}}$) are used to \res{tune} the reference distribution, then $n$ samples are generated for each $\beta_{k}$ from the power density $\mathcal{N}_{\beta_{k},j}(\mu_{\beta_{k},j},\,\sigma_{\beta_{k},j}^{2})$ with 
\begin{equation}
\begin{split}
   \mu_{\beta_{k},j} = \frac{\bar{\theta}_{j}(1-\beta_{k})}{S_{j}^{2}\;\beta_{k}\left(\frac{1+v}{v}\right)+(1-\beta_{k})}, \text{and}\\
   \sigma_{\beta_{k},j}^{2}= \left(\beta_{k}\left(\frac{1+v}{v}\right)+ \frac{(1-\beta_{k})}{S_{j}^{2}}\right)^{-1},
\end{split}
    \label{GSS analytic}
\end{equation}
where $\bar{\theta}_{j}$ and $S_{j}^{2}$ are respectively the mean and the variance of the calibration samples from the $j^{th}$ dimension.

\begin{figure}
         \centering
         \includegraphics[width=0.47\textwidth]{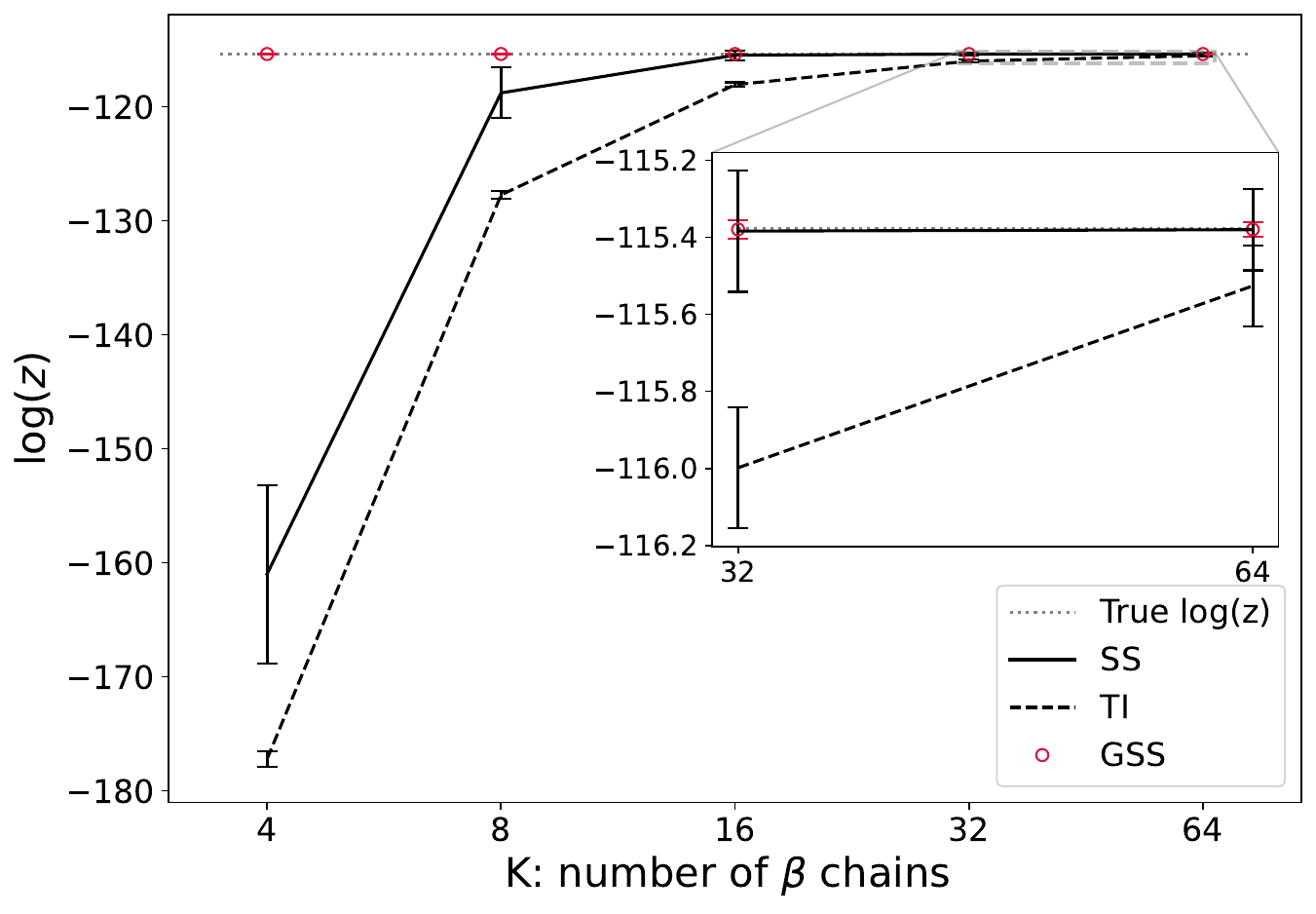}
         \caption{Comparison of log-marginal likelihood estimated as a function of $K$ number of $\beta$ chains for the Gaussian model with TI, SS, and GSS methods.}
         \label{fig:ss vs TI vs GSS}

\end{figure}

For this comparison, we set the number of dimensions to $d=50$ and variance to $v = 0.01$, resulting in a $\log(z)=-115.38$.  With $\beta_k$ uniformly dispersed along the quantiles of  $\text{Beta}(0.3,1)$, we fix $n=1000$ independent sample, $K\in[4,8,16,32,64]$ number of $\beta$ chains for each method. For the GSS estimator, the same setup is used, however, we set $N_{\text{cal}}=1000$ and $n=10$. To evaluate the uncertainty of each method's estimate, 1000 independent estimates are generated for each $K$. Figure~\ref{fig:ss vs TI vs GSS} shows that TI requires more than 32 chains and 16 chains for SS to converge to the true $\log(z)$, while the GSS estimator yields an accurate estimate of $\log(\widehat{z})=-115.37$ with $K=4$. This comparison demonstrates how the $\log(z)$ estimated using SS and TI can be biased even when error constraints are available. Thus, only increasing $K$ will ensure convergence and an accurate $\log(z)$ estimate. 

\begin{figure}
         \centering
         \includegraphics[width=0.47\textwidth]{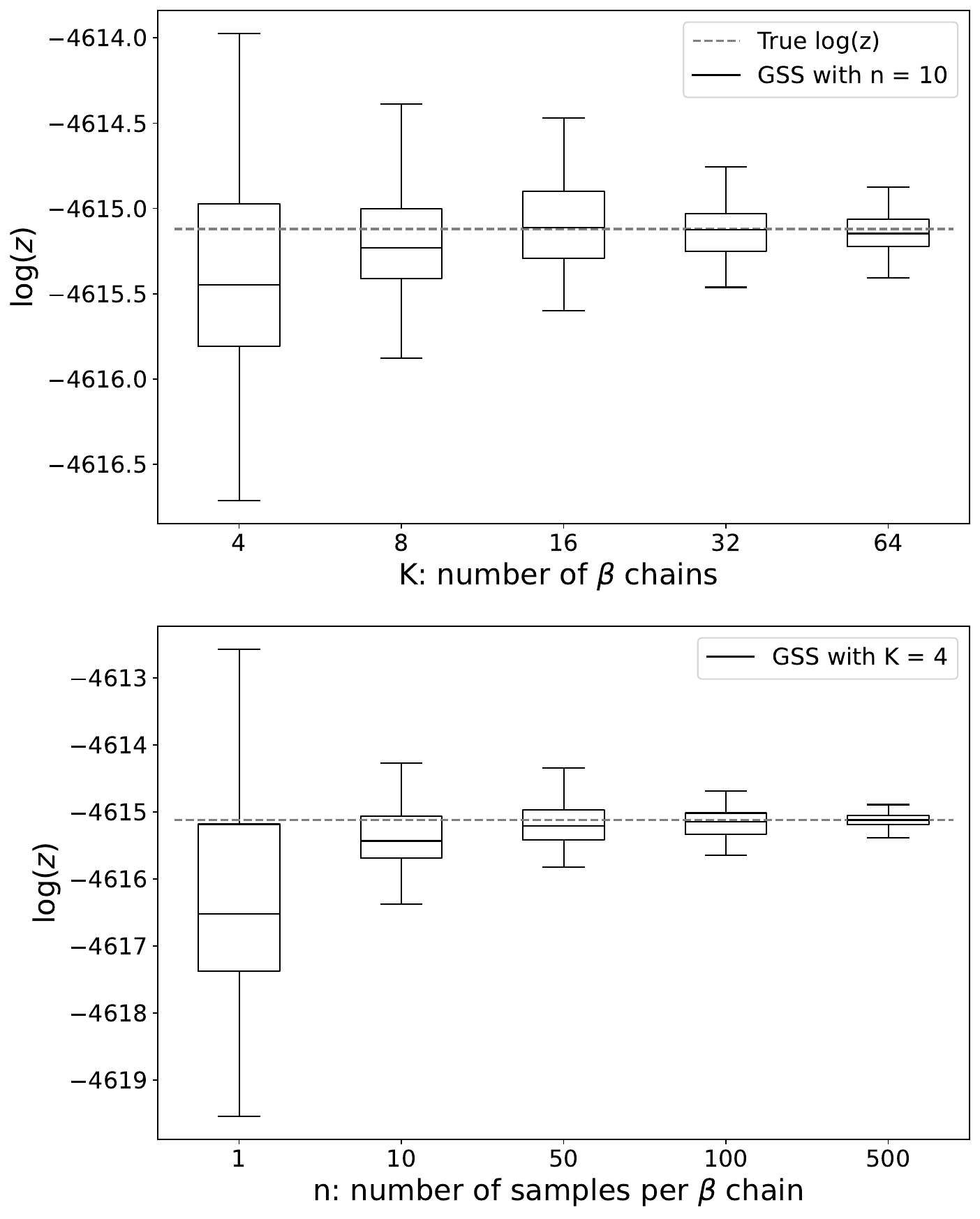}
         \caption{Convergence of the marginal likelihood estimated using GSS for the Gaussian model with $d=2000$ and $v=0.01$. Top panel:  Comparison of log-marginal likelihood estimated as a function of $K$ number of $\beta$ chains with $n=10$. Bottom panel: Comparison of log-marginal likelihood estimated as a function of $n$ number of independent samples per $\beta$ chains with $K=4$ . }
         \label{GSS 2000}

\end{figure}

The GSS estimator's performance remains consistent with a higher number of dimensions as shown in Figure~\ref{GSS 2000} where the number of dimensions $d = 2000$ and $N_{\text{cal}}=500$. Figure~\ref{GSS 2000} shows that one can either use fewer samples and increase $K$ or use few $\beta$ chains and increase $n$ the number of samples per chain. In practice, numerical experiments suggest that the computational cost should be allocated to the number of $\beta$ chains since a higher number of $K$ increases the similarity between the consecutive importance distributions \citep{russel2017bayesian}.


\section{Application of GSS to the PTA}
\label{Application to PTA}

In this section, we will illustrate how the GSS integrates into the PTA-Bayesian framework, and demonstrate the different benefits of estimating the \textit{marginal likelihood} in a PTA analysis. For this application, a Metropolis-Hastings algorithm \citep{chib1995understanding} was implemented to perform the sampling for the GSS method, where MPI for Python \citep{dalcin2005mpi} was employed to parallelize the sampling of $\beta$ chains. \res{This implementation was designed to adequately  fit into the Bayesian framework} used in PTA analyses \citep[$\text{ENTERPRISE}$;][]{ellis2019enterprise,enterprise}. In the following, the performance of GSS is assessed based on $K$, the number of $\beta$ chains, and $n_{\text{ESS}}$ the effective sample size (ESS) per  $\beta$ chain. We will use the GSS method to produce multiple independent estimates for each single analysis. These multiple estimates allow the direct comparison of all models by constructing a distribution of the Bayes factor of the compared models. In practice, single random draws of  $\log(\widehat{z})$ from the marginal likelihood distributions for model $M_1$ and $M_2$ are used to compute each $\log(\text{BF}_{M_{1}/M_{2}})$, the standard deviation of the resulting  $\log(\text{BF}_{M_{1}/M_{2}})$ distribution is used to measure uncertainty. 
%\pmr{what about the uncertainty? and direct BF calculation?}

%\pmr{I'd propose for the next 3 sentences: In the following, the performance of GSS is assessed based on the number of $\beta$ chains  $K$, and the effective sample size \citep[\textit{ESS};][]{martino2017effective}. We generate multiple estimates, allowing for the comparison of all models by constructing the distribution of Bayes factors for the compared models.}
\subsection{Application to NANOGrav}

This section demonstrates the application of the GSS to estimate the marginal likelihood for different models used in the PTA single pulsar noise analysis and GWB analysis. We reproduce the analysis and the public NANOGrav 15yr Data Release and the NANOGrav tutorials \citep{agazie2023nanograv} using $\text{ENTERPRISE}$. The consistency of the reproduced Bayesian analysis is compared with the results discussed in the NANOGrav 15 yr Data Set.
%\pmr{In table 1, should we not use something like $\log \bar{\widehat{z}}$ or $\bar{\log\widehat{z}}$?}

\begin{table*}    
    \centering
\caption{Mean and standard deviation of 100 independent log-marginal likelihood estimates for NANOGrav single pulsar noise models obtained via GSS with $K=8$.}
\label{tab:single noise log(z)}
\res{
    \begin{tabular}{|c|c|c|c|c|c|c|c|c|} \hline  
          &   \multicolumn{2}{|c|}{WN}&   \multicolumn{2}{|c|}{WN+DMGP}&  \multicolumn{2}{|c|}{WN+RN}& \multicolumn{2}{|c|}{WN+RN+DMGP}\\ \hline  
 PSR &  $\log(\widehat{z})$&std&  $\log(\widehat{z})$&std&  $\log(\widehat{z})$&std& $\log(\widehat{z})$&std\\ \hline  
         J1012+5307 &   275708.30&0.20&   275708.35&0.25&   275744.84&0.48&  275745.76&0.61\\ \hline  
         J1600-3053 &   261778.81&0.57&   261778.87&0.65&   261784.37&0.81&  261784.82&0.87\\ \hline  
         J1705-1903 &   111504.39&0.29&   111600.51&0.32&   111562.63&0.44&  111633.47&0.41\\ \hline  
         J1713+0747 &   750942.80&3.24&   750946.68&1.18&   751285.52&0.78&  751284.93&2.13\\ \hline 
    \end{tabular}    }
\end{table*}

\subsubsection{Model selection for single pulsar analysis}

The following four pulsars were selected from NANOGrav \citep{agazie2023nanograv}  as candidates for single-pulsar noise analysis: PSRs~J1012+5307, J1600-3053, J1705-1903, and J1713+0747. Around $100,000$ posterior samples are generated for each combination of the following noise models: white noise (WN), dispersion measure Gaussian process (DMGP), and red noise (RN). A 20,000 sample\res{s} of the burn-in period is discarded from the posterior samples \res{used} to calibrate the reference distribution $\pi_0$. In this case, the posterior samples are reasonably unimodal, therefore, for each model a normal distribution is used to approximate each parameter's marginal posterior distribution. The product of these distributions defines the reference distribution $\pi_0$, as it usually provides a sufficiently accurate approximation to the posterior  for the GSS method. The mean and the variance of each parameter's set of samples are used to calibrate $\pi_0$, then a mean effective sample size $n_{\text{ESS}}\approx10$ is generated for each $\beta$ chain with $K=8$  for each single estimate.  Table \ref{tab:single noise log(z)} reports the mean and standard deviation of 100 independent $\log(\widehat{z})$ for the combination of the three noise models. These values of $\log(\widehat{z})$ are used to compute the $\log(\widehat{\text{BF}})$ for comparing models' relative performance.

%\pmr{This can be confusing. I'd suggest: the WN model is used as a baseline and is compared to the best-performing model using the $\log(\text{BF})\pm1\sigma$}
For each pulsar, the WN model is used as a baseline for model comparison, the best-performing model compared to WN is reported with their $\log(\widehat{\text{BF}}_{\text{M/WN}})\pm1\sigma$. Across all four pulsars, strong statistical evidence is observed for the Red Noise model. For PSR~J1012+5307, WN+RN is favored with $\log(\widehat{\text{BF}})=36.47\pm0.56$. When combined with the other two models, the DMGP model has weak statistical evidence since $\log(\widehat{\text{BF}})<1$. For PSR~J1705-1903 and PSR~J1713+0747, the model comparison shows high statistical significance between each fitted model, where the most significant model is WN+RN+DMGP with $\log(\widehat{\text{BF}})=129.07\pm0.49$ for PSR~J1705-1903,  and WN+RN with  $\log(\widehat{\text{BF}})=342.58\pm3.29$ for PSR~J1713+0747. PSR~J1600-3053 demonstrates the least RN significance compared to the other 3 pulsars with $\log(\widehat{\text{BF}})=5.56\pm0.99$. The evidence of the excess noise becomes more significant for the model WN+RN+DMGP with $\log(\widehat{\text{BF}})=6.00\pm1.05$.
% \pmr{this is not precise}

One can further examine the performance of each model by inspecting the influence of its parameters $\boldsymbol{\delta}$ when included. This influence  can be demonstrated for the PTA models using the previous combinations of red noise models. This evaluation can be conducted using the \textit{Inclusion Bayes factor} \citep[IBF;][]{hinne2020conceptual} which examines the proportional change of the marginal likelihood when a set of parameters $\boldsymbol{\delta}$ are included in a set of models. Let $\mathcal{M}_{\boldsymbol{\delta}^1}$ and $\mathcal{M}_{\boldsymbol{\delta}^0}$  be respectively the sets of models that include and exclude $\boldsymbol{\delta}$ from their respective parameter vector. The IBF is given by
\begin{equation}
\text{IBF} = \frac{\sum_{M\in\mathcal{M}_{\boldsymbol{\delta}^1}} z(X|M)}{\sum_{M\in\mathcal{M}_{\boldsymbol{\delta}^0}} z(X|M)},
\label{IBF}
\end{equation}
%\begin{equation}
%\text{IBF} = %\frac{\sum_{M\in\mathcal{M}_{\boldsymbol{\delta}:1}} z(X|M)}{\sum_{M\in\mathcal{M}_{\boldsymbol{\delta}:0}} z(X|M)},
%\label{IBF1}
%\end{equation}
which denotes the sum of marginal likelihoods of models including $\boldsymbol{\delta}$ divided by the sum of marginal likelihoods of models excluding it. Table~\ref{tab:single ibf} shows the IBF to compare the overall performance of RN and DMGP parameters. IBF reinforces the strong  statistical evidence of RN across all pulsars while including DMGP parameters delivers a better performance only for PSR~J1705-1903 and PSR~J1713+0747. Since pulsar customized noise models are used for the full GWB analysis, the IBF can provide additional statistical evidence and help with a robust model selection where decisions are critical. 
\begin{table}
    \centering
    \begin{tabular}{|c|c|c|c|c|} \hline 
         &  \multicolumn{2}{|c|}{DMGP}&  \multicolumn{2}{|c|}{RN}\\ \hline 
         PSR&  $\log(\widehat{\text{IBF}})$&  std&  $\log(\widehat{\text{IBF}})$&std\\ \hline 
         J1012+5307 &  0.99&  0.87&  73.94&0.82\\ \hline 
         J1600-3053 &  0.52&  1.41&  11.42&1.43\\ \hline 
         J1705-1903 &  166.92&  0.75&  91.20&0.74\\ \hline 
         J1713+0747 &  3.46&  4.14&  680.88&4.17\\ \hline
    \end{tabular}
    \caption{Mean and standard deviation of 1000 independent $\log(\widehat{\text{IBF}})$ estimates for the parameter inclusion of DMGP model and RN model in NANOGrav single pulsar noise models.}
    \label{tab:single ibf}
\end{table}

\begin{figure}
         \centering
         \includegraphics[width=0.47\textwidth]{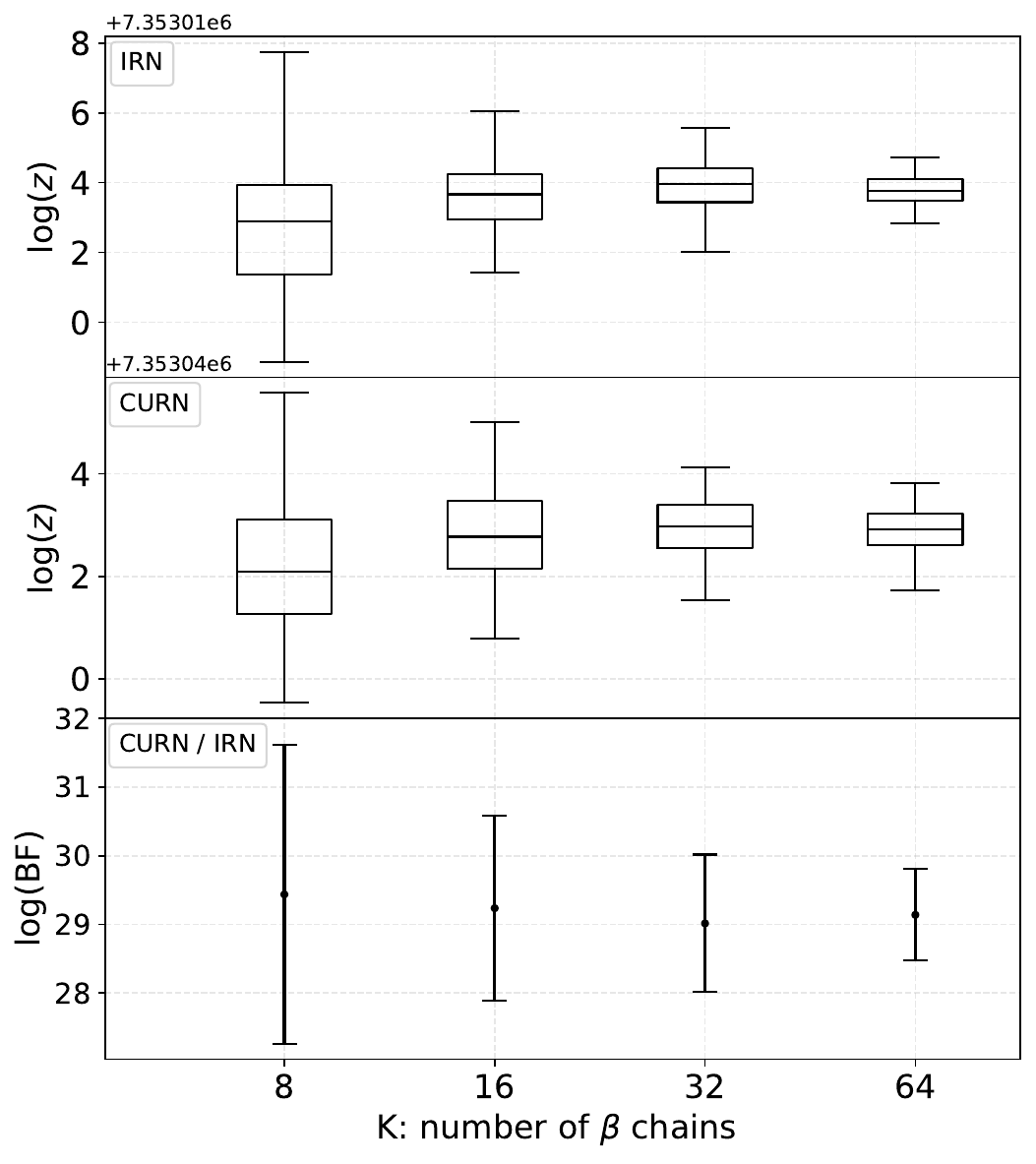}
         \caption{Estimations as a function of $K$ number of $\beta$ chains for the NANOGrav dataset via GSS. Top panel: log-marginal likelihood estimated for the IRN model. Middle panel: log-marginal likelihood estimated for the CURN model. Bottom panel: $\log(\widehat{\text{BF}}_\text{CURN/IRN})$ comparing CURN to IRN. }
         \label{fig:Nanograv}

\end{figure}
\subsubsection{Evidence estimation for the GWB analysis}
\label{GWB Nanograv}

% In this study, the Spl ORF uses 7 spline knots to pulsar angular separations located at $[10^{-3}, 25.0, 49.3, 82.5, 121.8, 150, 180]$ as free parameters to estimate the HD correlations at that angle.
In parallel with our previous analysis, the GSS estimator can also be used to investigate the evidence of GWB in the PTA. We reproduce the posterior \res{samples} used in the NANOGrav analysis where all CRS models share the log-spectral amplitude $\log(A_{M})$ and the spectral index $\gamma_{M}$ of the power law as free parameters for the four following models: intrinsic red noise (IRN), common uncorrelated red noise (CURN), spline overlap reduction function (Spl ORF), and the Hellings and Downs (HD) serving as the probe for GWB. The default setup and codes provided by the NANOGrav public data release are used in this demonstration. We refer to \cite{agazie2023nanograv} for extended details on the setup and the models. In this case, around $120,000$ samples are generated for each of the 4 models. Before estimating $\log(z)$, the posterior samples are compared with the previous analyses for consistency. The means and the standard deviations of the generated samples are in agreement with the reported results in \cite{agazie2023nanograv}. 

To configure the GSS, a 20,000 sample\res{s} of the burn-in period is discarded from the posterior samples used to \res{tune} the reference distribution $\pi_{0}$. By tuning the reference distribution and using it as a sampling proposal distribution, the GSS estimator enables us to skip the new burn-in period in the GSS sampling. In Figure~\ref{fig:Nanograv}, each $\log(z)$ estimate is computed using $n_{\text{ESS}}\approx50$ per each $\beta$ chain, while the number of parallel chains grows with $K\in[8,16,32,64]$. A higher number of $n_{\text{ESS}}$ is used to cross-check the consistency of the $\log(\widehat{z})$ estimates. As demonstrated in Figure~\ref{fig:Nanograv}, the GSS estimator produces a consistent median over $100$ $\log(\widehat{z})$ starting from $K=16$ while constraining the uncertainty as the number of chains increases, a higher number of $n_{\text{ESS}}$ can improve the estimate of the median $\log(\widehat{z})$  for  $K=8$ as shown in Figure~\ref{GSS 2000}. The $\log(\widehat{\text{BF}})$ is consistent through all $K$ and the slight variation would not affect the inference.

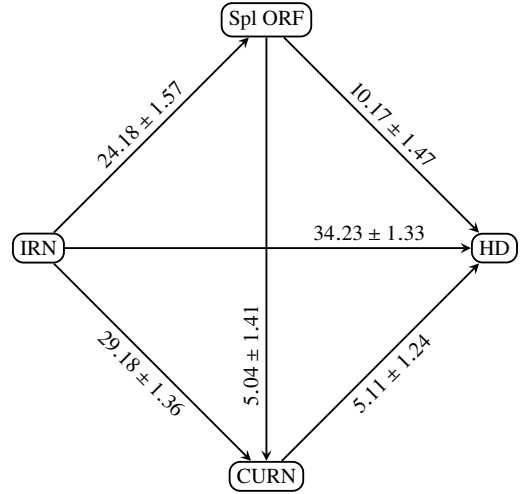
\begin{figure}
    \centering
    \begin{tikzpicture}
    % Nodes
    \node[draw, rounded corners, line width=0.7pt] (IRN) at (0,0) {IRN};
    \node[draw, rounded corners, line width=0.7pt] (Spl ORF) at (3,3) {Spl ORF};
    \node[draw, rounded corners, line width=0.7pt] (HD) at (6,0) {HD};
    \node[draw, rounded corners, line width=0.7pt] (CURN) at (3,-3) {CURN};

    % Arrows and labels
    \draw[->, >=stealth ,line width=0.7pt, scale=10] (IRN) to node[sloped, below] {$29.18 \pm 1.36$} (CURN);
    \draw[->,>=stealth , line width=0.7pt] (CURN) to node[sloped, below] {$5.11 \pm 1.24$} (HD);
    \draw[->,>=stealth , line width=0.7pt] (IRN) to node[sloped, above] {$24.18 \pm 1.57$} (Spl ORF);
    \draw[->,>=stealth , line width=0.7pt] (IRN) -- node[sloped, above, pos=0.75 ] {$34.23 \pm 1.33$} (HD);
    \draw[->,>=stealth , line width=0.7pt] (Spl ORF) to node[sloped, above, pos=0.75, rotate=180]  {$5.04 \pm 1.41$} (CURN);
    \draw[->,>=stealth , line width=0.7pt] (Spl ORF) to  node[sloped, above] {$10.17 \pm 1.47$} (HD);

    \end{tikzpicture}

    \caption{Mean and standard deviation of 1000 independent $\log(\widehat{\text{BF}}_{\text{M}_{1}/\text{M}_{2}})$ estimates comparing different common red process models in the NANOGrav dataset where the arrow starts at model $\text{M}_{2}$ and ends at model $\text{M}_{1}$. }
    \label{fig:bf diag}
\end{figure}

\begin{table}
    \centering
    \res{
    \begin{tabular}{|c|c|c|} \hline 
         &  $\log(\widehat{z})$& std\\ \hline 
         IRN&  7353013.52& 1.07\\ \hline 
         CURN&  7353042.72& 0.84\\ \hline 
         Spl ORF&  7353037.75& 1.18\\ \hline 
         HD&  7353047.80& 0.91\\ \hline
    \end{tabular}}
    \caption{Mean and standard deviation of 100 independent log-marginal likelihood estimates with $K=16$ for different common red process models in the NANOGrav dataset.}
    \label{tab:nanograv gwb log_z}
\end{table}
To evaluate the GWB evidence, 100 independent estimates were computed for each of the four models, Table~\ref{tab:nanograv gwb log_z} reports the mean and the standard deviation of  100 estimates of $\log(\widehat{z})$. Figure~\ref{fig:bf diag} reports the mean and standard deviation where 1000 estimates of $\log(\widehat{\text{BF}})$ were derived to compare the models. Compared to IRN, the other three models are highly supported reinforcing the strong evidence of an extra red process in the PTA dataset. Furthermore with a $\log(\widehat{\text{BF}}_\text{CURN/IRN})$ of $29.18 \pm 1.36$, a common uncorrelated red noise process is evident in the dataset reconfirming the previous PTA analyses \citep{arzoumanian2020nanograv,chen2021common,goncharov2022consistency}. Currently, CURN is favored over the Spl ORF model with $\log(\widehat{\text{BF}}_\text{CURN/Spl\;ORF}) = 5.04 \pm 1.41$. For the GWB evidence, $\log(\widehat{\text{BF}}_\text{HD/CURN}) = 5.11 \pm 1.24$ is consistent with the NANOGrav reported estimation of $ 5.42 \pm 4.25$, whereas the uncertainty is more constrained by the GSS method.
\begin{table*}
    
    \centering
\caption{Mean and standard deviation of 100 log-marginal likelihood estimates for each common red process model applied to the EPTA+InPTA datasets with K = 16.}
\label{tab:EPTA log(z)}
\res{
    \begin{tabular}{|c|c|c|c|c|c|c|c|c|} \hline  
           &\multicolumn{2}{|c|}{DR2new}&\multicolumn{2}{|c|}{DR2new+}&   \multicolumn{2}{|c|}{DR2full}&   \multicolumn{2}{|c|}{DR2full+}\\ \hline  
 &  $\log(\widehat{z})$&std&  $\log(\widehat{z})$&std&  $\log(\widehat{z})$&std& $\log(\widehat{z})$&std\\ \hline 
         PSRN+CURN &493684.31&1.02&525786.67&1.08&   607134.56&1.20&   639232.25&1.03\\ \hline  
         PSRN+Binned ORF  &493681.86&1.30&525783.84&1.36&   607130.75&1.31&   639228.67&1.05\\ \hline  
         PSRN+GWB &493692.55&1.14&525794.58&0.98&   607140.21&1.12&   639238.08&1.01\\ \hline
    \end{tabular}}

\end{table*}

\begin{table*}
    \centering
\caption{ Mean and standard deviation of 1000 independent $\log(\widehat{\text{BF}}_{\text{M/CURN}})$ estimates comparing Binned ORF and GWB model to CURN model across EPTA+InPTA datasets.}
\label{tab:EPTA BF}
    \begin{tabular}{|c|c|c|c|c|c|c|c|c|} \hline  
         Model&  \multicolumn{2}{|c|}{DR2new}&  \multicolumn{2}{|c|}{DR2new+}&  \multicolumn{2}{|c|}{DR2full}&  \multicolumn{2}{|c|}{DR2full+}\\ \hline 
 & $\log(\widehat{\text{BF}})$& std& $\log(\widehat{\text{BF}})$& std& $\log(\widehat{\text{BF}})$& std& $\log(\widehat{\text{BF}})$&std\\ \hline 
         PSRN+Binned ORF &   -2.35&1.65&   -2.81&1.76&   -3.62&1.42&   -3.80&1.81
\\ \hline  
 PSRN+GWB&  8.24&1.47&  7.89&1.48&  5.79&1.42&  5.70&1.57\\ \hline 
    \end{tabular}

\end{table*}
\subsection{Application to EPTA+InPTA}

\label{Application to EPTA+InPTA}
The EPTA+InPTA collaboration has publicly released its second dataset with the analysis for GW evidence \citep{antoniadis2023second}. This analysis combines the dataset of 25 pulsars into  four subsets: DR2full covers 24.7 years of EPTA data, DR2new is limited to the last 10.3 years of observations with the upgraded radio telescopes, DR2full+ and DR2new+ add 0.7 years of InPTA data to the previous subsets. The EPTA+InPTA study  uses customized pulsar noise models reported in \cite{antoniadis2023second2}. We will reproduce the analysis using the 4 subsets for the following models CURN, GWB (referred to as HD in section~\ref{GWB Nanograv}), and Binned ORF by using the default setup provided in the second data release. Around $120,000$ posterior samples are generated for each model, and then their means and standard deviations are examined for consistency with the reported ones in \cite{antoniadis2023second}. A burn-in period of the first 20,000 samples was discarded from the posterior samples used to calibrate the reference distribution $\pi_{0}$. A 100 $\log(\widehat{z})$ per model are estimated  with $K=16$ and $n_{\text{ESS}}\approx20$ samples per $\beta$ chain. To assess the stability of the estimation, $n_{ESS}$ and $K$ are increased and $\log(\widehat{z})$ estimates are cross-checked for consistency.
\begin{figure}
         \centering
         \includegraphics[width=0.48\textwidth]{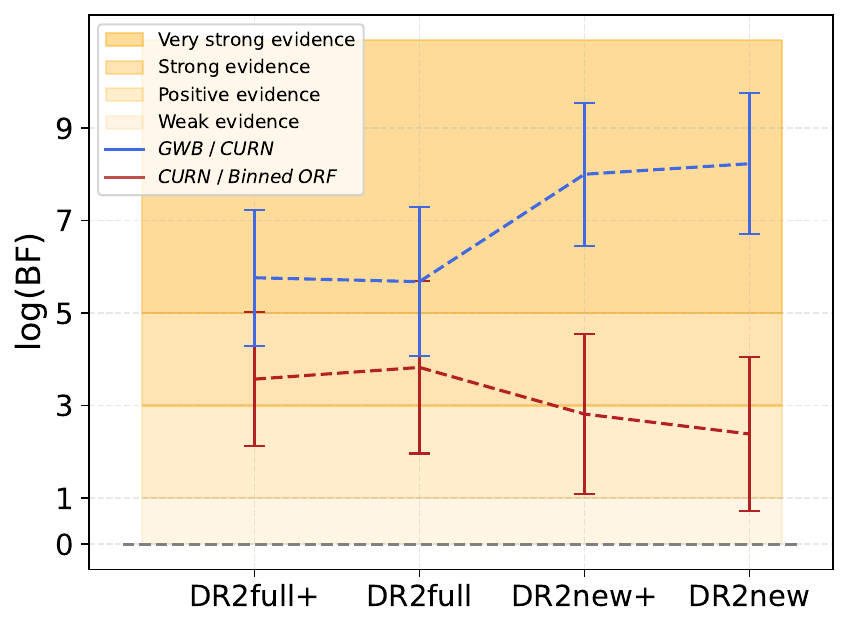}
         \caption{Evolution of $\log(\widehat{\text{BF}}_{\text{M}_{1}/\text{M}_{2}})$ comparing different common red process models as a function of the dataset size of  EPTA+InPTA dataset with DR2full+ the full dataset and DR2new the shortest subset. }
         \label{fig:EPTA evolve}

\end{figure}
Table~\ref{tab:EPTA log(z)} reports the mean and the standard deviation of 100 $\log(\widehat{z})$ estimates for the 3 models across all EPTA+InPTA datasets. To compare each model's performance, Table~\ref{tab:EPTA BF} reports the mean and the standard deviation of 1000  $\log(\text{BF}_{\text{M/CURN}})$ independent estimates of the other two models compared to the CURN model. For DR2full and DR2full+, the $\log(\text{BF}_{\text{GWB/CURN}})$ supports the GWB model over the CURN model. This support is further reinforced when fitting only to the new generation dataset DR2new. However, CURN model is favored over the Binned ORF model, and this evidence decreases for DR2new. The addition of the InPTA dataset in DR2full+ and DR2new+ slightly decreases the evidence on the GWB model and slightly increases the evidence favoring CURN over Binned ORF. Figure~\ref{fig:EPTA evolve} summarizes this evolution of evidence in the EPTA+InPTA dataset. Although the InPTA dataset is recent the slight decrease in evidence of GWB might arise from adding 1 year of dataset asymmetrically to only 12 pulsars out of 25 candidates which will directly affect the correlations between pulsars.

\section{Discussion}
\label{discussion}
In the simulation study, we highlighted the difference\res{s} between methods of power posterior and showed their relative performance. The GSS method has outperformed both TI and SS and proved efficient for higher dimensional cases. GSS can benefit from the previous efforts usually done in the parameter estimation phase. The success of a cheaper GSS estimation relies on having a good posterior approximation which shortens the path to an accurate estimate. Without such an approximation, the GSS estimator will require more effort (but still less than SS and TI) to achieve an accurate estimate. These $\log(z)$ estimates result in an accurate evaluation of the $\log(\text{BF})$ and robust interpretation of evidence supporting different hypotheses. This evidence can further be tested using the IBF criteria, which examines the influence of a particular set of parameters on the evidence.

Due to its \res{lower computational} cost, the GSS method enabled multiple $\log(z)$ estimates with high accuracy for each PTA-Bayesian analysis in this study. For single noise analysis, GSS has shown its ability to offer model selection without constraints, compared to the Hypermodel that can fail to provide evidence evaluation in cases such as estimation for RN cases \citep{smarra2023second}. In the NANOGrav dataset, the four pulsars have shown different levels of significance for RN and DMGP signals, where PSR~J1713+0747 showed extreme evidence for RN. Note that PSR~J1713+0747 contributed the most to the common red noise signal in the EPTA+InPTA dataset \citep{antoniadis2023second}. For NANOGrav GWB evidence, the GSS method has provided a consistent and more accurate estimate than the Hypermodel method of evidence supporting a GWB signal compared to a common red noise signal. Furthermore, GSS enabled a direct comparison of expensive models\res{,} efficiently eliminating the arbitrary weights required by the Hypermodel method.
The EPTA GWB analysis consistently reconfirms the evidence supporting a GWB signal in the PTA dataset. This evidence becomes more significant for the new-generation dataset as shown in Figure~\ref{fig:EPTA evolve}. 

The PTA experiments rely on model selection to tune each part of the Bayesian noise model which increases the number of Bayesian analyses required to achieve better sensitivity. The current GSS setup substantially minimize\res{s} the cost of a single $\log(z)$ estimate for both light and expensive models, and would further benefit in the future from better samplers.  The efficiency of the GSS method will help reduce the cost of evidence estimation for the continuously growing model portfolio for GWs and new physics in the PTA \citep{afzal2023nanograv,smarra2023second}.

\section*{Acknowledgments}

EMZ, PMR, WS, and RM gratefully acknowledge support by the Marsden Fund Council grant MFP-UOA2131 from New Zealand Government funding, managed by the Royal Society Te Apārangi. 
This work was performed on the OzSTAR national facility at Swinburne University of Technology. The OzSTAR program receives funding in part from the Astronomy National Collaborative Research Infrastructure Strategy (NCRIS) allocation provided by the Australian Government, and from the Victorian Higher Education State Investment Fund (VHESIF) provided by the Victorian Government.

\vspace{5pt}\textit{Software}: \texttt{ENTERPRISE} \citep{ellis2019enterprise,enterprise}, \texttt{enterprise\_extension} \citep{enterprise}, \texttt{PTMCMC} \citep{justin_ellis_2017_1037579}  , \texttt{MPI4PY} \citep{dalcin2005mpi}, \texttt{Jupyter} \citep{kluyver2016jupyter}, \texttt{matplotlib} \citep{hunter2007matplotlib}, \texttt{numpy} \citep{harris2020array}, \texttt{scipy} \citep{pauli_virtanen_2020_4406806}, \texttt{arviz} \citep{kumar2019arviz}.

%%%%%%%%%%%%%%%%%%%%%%%%%%%%%%%%%%%%%%%%%%%%%%%%%%
\section*{Data Availability}
Data used in this analysis are available on GSS-estimator GitHub \footnote{https://github.com/nz-gravity/GSS-estimator}.


%\pmr{Check capital letters in the references, for instance, ``american statistician", ``Systematic biology". I'd also add the DOI to each reference}

\bibliographystyle{mnras}
\bibliography{biblio} % if your bibtex file is called example.bib

%\begin{thebibliography}{99}
%\end{thebibliography}

%%%%%%%%%%%%%%%%%%%%%%%%%%%%%%%%%%%%%%%%%%%%%%%%%%

%%%%%%%%%%%%%%%%% APPENDICES %%%%%%%%%%%%%%%%%%%%%
%%%%%%%%%%%%%%%%%%%%

% Don't change these lines
\bsp	% typesetting comment
\label{lastpage}
\end{document}